\documentclass[superscriptaddress,twocolumn,showpacs,pre]{revtex4}
\newcommand{\figurewidth}{\columnwidth}
\newcommand{\av}{_{\rm av}}

\usepackage{amsmath,graphicx,tabularx}
\usepackage{graphicx,amsmath}

\begin{document}

\title{
Phase Transitions in the 1-d Long-Range Diluted Heisenberg Spin Glass}

\author{Auditya Sharma}
\affiliation{Department of Physics, University of California,
Santa Cruz, California 95064}
\author{A.~P.~Young}
\email{peter@physics.ucsc.edu}
\affiliation{Department of Physics, University of California,
Santa Cruz, California 95064}

\date{\today}

\begin{abstract}
We use Monte Carlo simulations to
study the one-dimensional
long-range diluted Heisenberg spin glass with interactions
that fall as a power, $\sigma$, of the distance.  Varying the power is argued
to be equivalent to varying the space dimension of a short-range model.
We are therefore able to study both the mean-field and non-mean-field regimes.
For one value of $\sigma$, in the non-mean-field regime,
we find evidence that the chiral glass transition
temperature may be somewhat higher than the spin glass transition temperature. For
the other values of $\sigma$ we see no evidence for this.
\end{abstract}
%\pacs{03.67.Lx , 03.67.Ac, 64.70.Tg,75.10.Nr}
\pacs{75.50.Lk, 75.40.Mg, 05.50.+q}
\maketitle

\section{Introduction}
\label{sec:intro}
In the theory of phase transitions it is often helpful to study models in a
range of dimensions ranging from above the ``upper critical dimension'',
$d_u$, where
mean-field critical behavior is expected, to below the ``lower critical
dimension'', $d_l$, where fluctuations destroy the transition. However, it has been
difficult to do this numerically for spin glasses, since $d_u = 6$ is 
quite large, and slow dynamics
prevents more than a few thousand spins being equilibrated at low temperature
$T$. It follows that, at and above $d_u$,
one cannot study a sufficient range of sizes to perform the necessary
finite-size scaling analysis. 

As a result, there has been a lot of recent attention on long-range models in
one-dimension, in which the interactions fall off with a power of the
distance. Such models have a venerable history going back to 
Dyson~\cite{dyson:69,dyson:71}, who considered a ferromagnet with
interactions falling off like $1/r^\sigma$, and found a
paramagnet-ferromagnet transition for $1 < \sigma
\leq 2$. Kotliar et al.~\cite{kotliar:83} studied the spin glass version of
this model, which has received a lot of attention numerically in the last few
years~\cite{katzgraber:05,katzgraber:09,leuzzi:08,moore:10,leuzzi:09}.
%The
%main line of attack has relied on help from Monte Carlo methods, where a
%crucial point of contention is the system size that maybe studied. In a bid to
%stretch on the system-size-front, Leuzzi et al.~\cite{leuzzi:08} introduced
%the elegant diluted version of this model to reduce the complexity of the
%algorithm, and thus reach for bigger system sizes. 

There are few analytical results on spin glasses beyond mean field theory. For
the long-range models, Kotliar et al.~\cite{kotliar:83} computed critical
exponents in an expansion away
from the point where mean field theory occurs ($\epsilon$-expansion),
but, as we shall see, this is
poorly converged. Hence, most of what we know has come from numerical work.
Much of this, including numerics on long-range spin glass models, has
studied the Ising version. However, there are also reasons to study models with
vector spins, such as the Heisenberg (3-component) model.

One motivation is
that Kawamura~\cite{kawamura:87} proposed that there are two separate
transitions in vector spin glasses, a spin glass transition at $T=T_{SG}$ and
a ``chiral glass transition'' at higher temperature
$T_{CG}$, involving a freezing of vortex-like variables called chiralities.
While the original scenario had
$T_{SG} = 0$, it now appears that $T_{SG}$ is non-zero in three or more
dimensions, but the question of whether
$T_{SG} < T_{CG}$, or whether there is a single transition at which both types
of ordering occur, is still
open~\cite{hukushima:00,hukushima:05,matsubara:01,endoh:01,nakamura:02,picco:05,lee:03,lee:07,campos:06,fernandez:09b,viet:10}.
%Heisenberg spin glasses( and X-Y spin glasses) is driven not by the ordering
%of the spins, but rather by the ordering of three-spin, scalar-variables
%called chiralities. It has been a long-standing argument of Kawamura and
%collaborators that Heisenberg spin glasses exhibit chiral-glass ordering at a
%temperature that is higher than the spin-glass ordering temperature. This has
%been a controversial issue, with several works reporting evidence both for,
%and against the thesis.

A second motivation for studying the Heisenberg, rather than Ising, spin glass
is that it is possible to study larger sizes, see for example
Ref.~\cite{fernandez:09b}, which should be helpful in a finite-size scaling
analysis. In a second paper in this series~\cite{sharma:11b}, we will
investigate whether there is a de Almeida-Thouless~\cite{almeida:78} (AT) line
of transitions in a magnetic field for
Heisenberg spin glasses. This follows our recent work~\cite{sharma:10} which
shows that there \textit{is} an AT line for vector spin glasses provided one
considers a random field. The ability to study larger sizes will be
particularly useful for the AT-line study.
%Here, we present data only for zero field.

%In this study, we consider the  model of Leuzzi et al.~\cite{leuzzi:08} with
%Heisenberg spins, the undiluted version of which has been studied by Viet and
%Kawamura~\cite{viet:10} recently. Having worked out in a recent paper that
%arbitrary $m$ - component vector spin glasses also display a de
%Almeida-Thouless line in mean-field theory~\cite{sharma:10}, the present
%zero-field study on the diluted Heisenberg spin glass model is an essential
%step enroute our journey to address the question of whether or not an AT line
%exists for this model. In this work, we perform large-scale Monte Carlo
%simulations on system sizes larger than considered before, and follow this
%with extensive and careful data-analysis to identify the phase-transitions of
%this model.

Here we present data for the zero field transition for the Heisenberg spin
glass for values of the parameter $\sigma$ corresponding to (i) the mean-field
regime, (ii) the non-mean-field regime, and (iii) the borderline case where
the transition disappears. Most of our results find no evidence for separate
spin-glass and chiral-glass transitions. However, for one set of parameters in
the non-mean-field regime, the data indicates that $T_{CG}$ may be
somewhat greater than $T_{SG}$.
Whether this result remains valid in the thermodynamic limit, will require
future studies on significantly larger systems.

The plan of this paper is as follows. In Sec.~\ref{sec:model} we describe the
model that we study, while in Sec.~\ref{sec:numerics} we gives some technical details
of the simulations. The results are presented in Sec.~\ref{sec:results}
and our conclusions summarized in Sec.~\ref{sec:conclusions}.

\section{Model}
\label{sec:model}
We consider the Hamiltonian
\begin{equation}
\mathcal{H} = -\sum_{\langle i, j \rangle} J_{ij} \mathbf{S}_i \cdot \mathbf{S}_j,
\label{Ham}
\end{equation}
where $\mathbf{S}_i$ are classical $3$-component Heisenberg spins of length
$1$, and the interactions
$J_{ij}$ are independent variables with zero mean and a variance 
which falls off with a power of the distance between the
spins,
\begin{equation}
[J_{ij}^{2}]_{av} \propto \frac{1}{r_{ij}^{2\sigma}}, 
\label{J2}
\end{equation} 
where $[\cdots]_{av}$ means an average over disorder. In the version used in
early studies~\cite{katzgraber:05},
every spin interacts with every other spin with a strength which falls off, on
average, like Eq.~\eqref{J2}. However, this means that the CPU time per sweep
varies as $N^2$, rather than $N$, so large sizes can not be studied.
This problem was solved by Leuzzi et. al.~\cite{leuzzi:08} who proposed a
model in which,
instead of the \textit{magnitude} of the interaction falling off with distance like
Eq.~\eqref{J2}, it is the \emph{probability} of
there being a non-zero interaction between sites $(i,j)$ which falls off,
and when an
interaction does exist, its 
variance is independent of $r_{ij}$. The mean
number of non-zero interactions from a site, which we call $z$,
can be fixed, and we 
take $z = 6$.
To generate the set of pairs $(i,j)$ that have an interaction with the desired
probability
%is converted into: \begin{equation} P[J_{ij} \neq 0] \propto
%\frac{1}{r_{ij}^{2\sigma}}.  \end{equation} This has the advantage of making
%the energy $O(N)$ as opposed to $O(N^2)$, and therefore allows the simulation
%of larger size lattices. To generate bonds satisfying this condition,
%we start by fixing the total number of bonds in the $N$ site 
%system as $Nb=N*z/2$, where $z \equiv 6$ is the average coordination number. In contrast to Leuzzi et al~\cite{leuzzi:08}, 
%we prefer using the geometric distance between sites $i,j$ is defined as:
%\begin{equation}
 %r_{ij}=\frac{N}{\pi}sin[\frac{\pi}{N}(i-j)].
%\end{equation}
%Following Leuzzi et al~\cite{leuzzi:08}, we produce these bonds as follows. 
we choose spin $i$ randomly, and then choose
$j \ (\ne i)$ at distance $r_{ij}$ with probability 
\begin{equation}
p_{ij} = \frac{r_{ij}^{-2\sigma}}{\sum_{j\, (j\neq i)}r_{ij}^{-2\sigma}} \, ,
\label{pij}
\end{equation}
where, for $r_{ij}$, we put the sites on a circle and use the distance of the
chord, i.e.
\begin{equation}
r_{ij}=\frac{N}{\pi}\sin\left[\frac{\pi}{N}(i-j)\right].
\end{equation}
If $i$ and $j$ are already connected, we repeat the process until we find a pair
which has not been connected before. We then connect $i$ and $j$ with an
interaction picked from a Gaussian interaction
whose mean is zero and whose standard deviation is $J$, which set equal to 1.
This process is
repeated precisely $N_b = z N / 2 $ times. 

The result is that each pair $(i, j)$ will be
connected with a probability $P_{ij}$ which must satisfy the condition
$N \sum_j P_{ij} = N z$ since $P_{ij}$ only depends on $|i-j|$, $P_{ii}=0$, and
there are precisely $N z /2 $ connected pairs. 
It follows that, for a fixed site $i$,
\begin{equation}
\sum_j [\,J_{ij}^{2}\,]_{av} = J^2 \sum_j P_{ij} = J^2 z \, .
\label{sumJij2}
\end{equation}
Note that $P_{ij}$ is different from $z \times p_{ij}$ in Eq.~\eqref{pij} because
of the constraint
that no bond can occur twice. 
The mean-field spin glass
transition temperature for $m$-component vector spins is given
by~\cite{almeida:78b}
\begin{equation}
T_{SG}^{MF} = \frac{1}{m} \left(\sum_j [\,J_{ij}^{2}\,]_{av}\right)^{1/2} 
= {\sqrt{z} \over m} \, J \, ,
\label{TcMF}
\end{equation}
where the last equality follows from Eq.~\eqref{sumJij2}.
We set $J=1$ so that, for the situation here,
\begin{equation}
J = 1,\ z=6,\ m=3\, ,
\end{equation}
we have
\begin{equation}
T_{SG}^{MF} = {\sqrt{6} \over 3} \simeq 0.816\, ,
\end{equation}
the same as for the nearest-neighbor Heisenberg spin glass on a simple cubic
lattice.

By varying $\sigma$ one finds different types of
behavior~\cite{kotliar:83,katzgraber:03}, as shown in Table~\ref{tab:ranges}.
For $\sigma < 1/2$ the model is non-extensive (for instance the mean-field
transition temperature in Eq.~\eqref{TcMF} diverges) unless the interactions are
scaled by an inverse power of the system size. We will call this ``infinite
range''. The extreme limit of this, $\sigma=0$, is the
Sherrington-Kirkpatrick~\cite{sherrington:75} model, whose exact solution was
found by Parisi~\cite{parisi:79,parisi:80,parisi:83}. In fact, it has been
suggested~\cite{mori:11} (see also Ref.~\onlinecite{mori:10}) that, in the
thermodynamic limit, the behavior of the model
is \textit{identical} to that of the SK model for the \textit{whole range}
$0 \le \sigma < 1/2$.

The model is extensive for $\sigma > 1/2$ and a finite temperature transition
is expected for $\sigma < \sigma_l$, where the ``lower critical'' value is
\begin{equation}
\sigma_l = 1\, .
\end{equation}
The transition is in the mean-field universality
class~\cite{kotliar:83} for $\sigma < \sigma_u$, where the ``upper
critical'' value is
\begin{equation}
\sigma_u = 2/3\, . 
\end{equation}
For $\sigma_u < \sigma < \sigma_l$,
there is a finite-temperature transition with non mean-field
critical exponents.
In this paper we will study both
mean-field and non mean-field regions. Finally for $\sigma > \sigma_l$ the
there is no transition at finite temperature.

\begin{table}[!tb]
\caption{
A summary of the behavior for different ranges of $\sigma$ in one space
dimension and at zero field.
Infinite
range means that $\sum_{j\, (j\ne i)} J_{ij}^2$ diverges unless the bonds $J_{ij}$
are scaled by an inverse power of the system size.
%LR means that the behavior
%is dominated by the long-range tail of the interactions, and SR means that
%the behavior is that of a short-range system.
The behavior is mean-field like
for $\sigma < \sigma_u$ where $\sigma_u = 2/3$, and a finite-temperature
transition no longer occurs for $\sigma > \sigma_l$ where $\sigma_l = 1$.
\label{tab:ranges}
}
\begin{tabular*}{\columnwidth}{@{\extracolsep{\fill}} l l }
\hline
\hline
$\sigma$ & Behavior  \\
\hline
$\sigma = 0$ & SK model \\
$0 < \sigma \le 1/2$ & Infinite range  \\
%$1/2 < \sigma < 2/3$ & LR (mean field with $T_{\rm SG} > 0$) \\
%$2/3 < \sigma \le 1$ & LR (non-mean field with $T_{\rm SG} > 0$) \\
%$1 < \sigma \le 2$ & LR ($T_{\rm c} = 0$) \\
%$\sigma \ge 2$ & SR ($T_{\rm c} = 0$) \\
$1/2 < \sigma < 2/3$ & Mean-field with $T_{\rm SG} > 0$ \\
$2/3 < \sigma < 1$ & Non-mean-field with $T_{\rm SG} > 0$ \\
$1 < \sigma $ & $T_{\rm SG} = 0$ \\
\hline
\hline
\end{tabular*}
\end{table}

\section{Numerical Setup}
\label{sec:numerics}
We perform large scale Monte-Carlo simulations for 
$\sigma = 0.6,0.75,0.85$ and $1$. From the previous section we note that
$\sigma = 0.60$ is in the mean-field regime, 
$\sigma = 0.75$ and $0.85$
are in the non mean-field regime, and $\sigma = 1$ is the borderline case,
$\sigma = \sigma_l$,
beyond which there is no transition. A plausible scenario is that 
$T_{SG}=0$ for $\sigma = 1$, though the possibility that $T_{SG}$ is non-zero
cannot be ruled out \textit{a priori}. 
Table \ref{simparams} lists the
parameters of the simulation.

\begin{table}
\caption{
Parameters of the simulations.  $N_{\rm samp}$ is the number of samples,
$N_{\rm equil}$ is the number of overrelaxation Monte Carlo sweeps for
equilibration for each of the $2 N_T$ replicas for a single sample. The same
number of sweeps is done in the measurement phase, with a measurement
performed every four overrelaxation sweeps.
%$N_{\rm meas}$ is the number of overrelaxation sweeps for measurement.
The number of heatbath sweeps is equal to
10\% of the number of
overrelaxation sweeps.  $T_{\rm min}$ and $T_{\rm max}$ are the
lowest and highest temperatures simulated,
and $N_T$ is the number of temperatures
used in the parallel tempering.
\label{simparams}
}
\begin{tabular*}{\columnwidth}{@{\extracolsep{\fill}}r r r r r r r r }
\hline
\hline
$\sigma$ & $N$  & $N_{\rm samp} $ & $N_{\rm equil}$ & $T_{\rm min}$ &
$T_{\rm max}$ & $N_{T}$  \\
\hline
$0.6$ &   $128$ &  $16000$ &   $128$ & $0.20$ & $0.70$ &  $40$ \\
$0.6$ &   $256$ &  $16000$ &   $256$ & $0.20$ & $0.70$ &  $40$ \\
$0.6$ &   $512$ &  $16000$ &   $512$ & $0.20$ & $0.70$ &  $40$ \\
$0.6$ &  $1024$ &  $16000$ &  $1024$ & $0.20$ & $0.70$ &  $40$ \\
$0.6$ &  $2048$ &  $16000$ &  $2048$ & $0.20$ & $0.70$ &  $40$ \\
$0.6$ &  $4096$ &   $6100$ &  $4096$ & $0.30$ & $0.70$ &  $40$ \\
$0.6$ &  $8192$ &   $1000$ &  $8192$ & $0.30$ & $0.70$ &  $50$ \\
$0.6$ & $16384$ &    $500$ & $16384$ & $0.35$ & $0.70$ &  $55$ \\
$0.6$ & $32768$ &    $400$ & $32768$ & $0.35$ & $0.70$ &  $60$ \\
\hline
$0.75$ &   $128$ &  $8000$ &   $128$ & $0.20$ & $0.55$ &  $40$ \\
$0.75$ &   $256$ &  $8000$ &   $256$ & $0.20$ & $0.55$ &  $40$ \\
$0.75$ &   $512$ &  $8000$ &   $512$ & $0.20$ & $0.55$ &  $40$ \\
$0.75$ &  $1024$ &  $3000$ &  $1024$ & $0.20$ & $0.55$ &  $40$ \\
$0.75$ &  $2048$ &  $3000$ &  $2048$ & $0.20$ & $0.55$ &  $40$ \\
$0.75$ &  $4096$ &  $3000$ &  $4096$ & $0.20$ & $0.55$ &  $40$ \\
$0.75$ &  $8192$ &  $1100$ &  $8192$ & $0.20$ & $0.55$ &  $50$ \\
$0.75$ & $16384$ &   $500$ & $16384$ & $0.25$ & $0.55$ &  $55$ \\
$0.75$ & $32768$ &   $400$ & $32768$ & $0.25$ & $0.55$ &  $60$ \\
\hline
$0.85$ &   $128$ &  $16000$ &    $512$ & $0.09$ & $0.30$ &  $40$ \\
$0.85$ &   $256$ &  $16000$ &   $1024$ & $0.09$ & $0.30$ &  $40$ \\
$0.85$ &   $512$ &  $10000$ &   $2048$ & $0.09$ & $0.30$ &  $40$ \\
$0.85$ &  $1024$ &  $10000$ &   $8192$ & $0.14$ & $0.22$ &  $20$ \\
$0.85$ &  $2048$ &   $8000$ &  $16384$ & $0.14$ & $0.22$ &  $20$ \\
$0.85$ &  $4096$ &   $4000$ &  $32768$ & $0.14$ & $0.22$ &  $36$ \\
$0.85$ &  $8192$ &   $2000$ &  $65536$ & $0.15$ & $0.21$ &  $20$ \\
$0.85$ & $16384$ &   $1700$ & $131072$ & $0.16$ & $0.21$ &  $20$ \\
\hline
$1.0$ &   $128$ &  $2000$ &     $2048$ &  $0.03$ & $0.10$ &  $10$ \\
$1.0$ &   $256$ &  $2000$ &     $4096$ &  $0.03$ & $0.10$ &  $10$ \\
$1.0$ &   $512$ &  $2000$ &    $16384$ &  $0.02$ & $0.10$ &  $40$ \\
$1.0$ &  $1024$ &  $1200$ &   $524288$ & $0.017$ & $0.08$ &  $40$ \\
$1.0$ &  $2048$ &   $500$ &  $2097152$ & $0.017$ & $0.08$ &  $60$ \\
\hline
\hline
\end{tabular*}
\end{table}

\subsection{Equilibration}
As discussed in earlier work~\cite{katzgraber:01,katzgraber:09}
there is a convenient test for
equilibration with Gaussian interactions, namely the relationship
\begin{equation}
U = {J^2 \over T} \, {z \over 2} \, (q_l - q_s) \, ,
\label{equiltest}
\end{equation}
is valid in equilibrium but the two sides approach their common equilibrium
value from opposite directions as equilibrium is approached. Here
$U = -(1/N)[\sum_{\langle i, j \rangle} \epsilon_{ij} J_{ij} \langle {\bf S}_i \cdot
{\bf S}_j \rangle ]\av $ is
the average energy per spin,
$q_l = (1/N_{b})\sum_{\langle i, j \rangle} \epsilon_{ij} 
[ \langle 
{\bf S}_i \cdot {\bf S}_j \rangle^2]\av$ is the ``link overlap'', and
$q_s = (1/N_{b})\sum_{\langle i, j \rangle} \epsilon_{ij}
[\langle ({\bf S}_i \cdot {\bf S}_j)^2
\rangle]\av$, where $N_{b}=z N/2$,
and $\epsilon_{ij} = 1$ if there is a bond between $i$ and $j$ and is zero
otherwise.
Equation~(\ref{equiltest}) is easily derived by integrating by parts the
expression for the average energy with respect to $J_{ij}$ since it has
a Gaussian distribution. Note that in the numerics we set $J = 1$.

We determine both sides of 
Eq.~\eqref{equiltest} for different numbers
of Monte Carlo sweeps (MCS) which increase
in a logarithmic manner, each value being twice the previous one. In all
cases we 
average over the last half of the sweeps. We consider the data to be
equilibrated, if, when averaging over a large number of samples,
Eq.~\eqref{equiltest} is satisfied for at least the last two points. 

\subsection{Simulation Technology}
To equilibrate the system in as small a number of sweeps as
possible, with the minimum amount of CPU time, we perform three types of
Monte Carlo sweeps~\cite{lee:07,campos:06,fernandez:09b}.

%\begin{enumerate}
%\item
The workhouse of our simulation is the ``Microcanonical'' sweep~\cite{alonso:96}
(also known as an ``over-relaxation'' sweep).
We sweep sequentially through the lattice, and, at
each site, compute the local field on the spin, $\mathbf{H}_i = \sum_j J_{ij}
\mathbf{S}_j$. The new value for the spin on site $i$ is taken to be
its old value reflected about $\mathbf{H}$, i.e.
\begin{equation}
\mathbf{S}'_i = -\mathbf{S}_i + 2\, {\mathbf{S}_i \cdot \mathbf{H}_i \over
H_i^2}\, \mathbf{H}_i \, .
\label{reflect}
\end{equation}
These sweeps are microcanonical
because they preserve energy. They are very fast because the
operations are simple and no random numbers are needed.  For reasons
that are not fully understood, it also seems that they ``stir up'' the
spin configuration very efficiently~\cite{campos:06} and the system
equilibrates faster than if one only uses ``heatbath'' updates,
described next, see e.g.~Fig.~9 of Ref.~[\onlinecite{pixley:08}].

We also need to do heatbath sweeps in order to change the energy.
As for the
microcanonical case, we sweep sequentially through the lattice.
We take the direction of the local field $\mathbf{H}_i$,
to be the
polar axis for the spin on
site $i$. We compute the polar and azimuthal angle of the new spin
direction relative
to the local field by the requirement that this direction occurs with the
Boltzmann probability, see Ref.~[\onlinecite{lee:07}] for details.

Finally we perform parallel tempering sweeps~\cite{hukushima:96,marinari:98b}
to prevent the system from being trapped in local minima at low temperature.
We take $N_T$ copies of the system with the same bonds but at
a range of
different temperatures.
The minimum temperature, $T_{\rm min} \equiv T_1$,
is the low temperature where one
wants to investigate the system (below $T_{SG}$ in our case), and the
maximum,
$T_{\rm max} \equiv T_{N_T}$,
is high enough that the
the system equilibrates very fast
(well above $T_{SG}$ in our case). A parallel tempering sweep consists
of swapping the temperatures of the spin configurations at a pair of
neighboring temperatures, $T_i$ and $T_{i+1}$, for $i = 1, 2, \cdots ,
T_{N_T - 1}$ with a probability that satisfies the detailed balance
condition. The Metropolis probability for this is~\cite{hukushima:96}
\begin{equation}
\quad P(T\ \mbox{swap}) = \left\{
\begin{array}{ll}
\exp(\Delta \beta \, \Delta E),  & (\mbox{if} \ \Delta \beta \, \Delta E
< 0), \\
1, & (\mbox{otherwise}), 
\end{array}
\right.
\label{PTswap}
\end{equation}
where $\Delta \beta= 1/T_i - 1/T_{i+1}$ and $\Delta E = E_i - E_{i+1}$,
in which $E_i$ is the energy of the copy at temperature $T_i$. In this
way, a given set of spins (i.e.~a copy)
performs a random walk in temperature space.

%To facilitate equilibration, we perform three kinds of Monte-Carlo
%sweeps~\cite{lee:07,campos:06,fernandez:09b}:
%``overrelaxation", ``heatbath" and parallel tempering,
%Of these, overrelaxation sweeps (also called
%microcanonical sweeps because they conserve energy) are the workhorse because
%they are fast and require no random numbers. Heatbath sweeps, which change the
%energy are essential to ensure ergodicity; however, they involve somewhat
%complicated operations, and therefore, we perform typically one heatbath sweep
%for every ten overrelaxation sweeps. Surprisingly, a combination of mainly
%overrelaxation sweeps with some heatbath sweeps, equilibrates the system 
%in fewer total sweeps than if all the sweeps are heatbath, see
%Fig.~9 of Ref.~[\onlinecite{pixley:08}] for data on 
%the similar situation in XY spin glasses.
%
%Parallel tempering is a
%standard technique to ensure that the system does not get caught in local
%minima at low temperatures. It involves simulating several copies of the
%system at various temperatures at once and swapping spin configurations at
%neighboring temperatures with a probability which satisfied the detailed
%balance condition.

We perform one parallel tempering sweep for
every ten overrelaxation sweeps. Since there are two copies of spins
\textit{at each
temperature}, indicated by labels ``$(1)$'' and ``$(2)$'' in Eq.~\eqref{qmunu}
below, we actually perform parallel tempering sweeps among the set of $N_T$
copies labeled ``$(1)$'' and, separately, among the set of  $N_T$ copies
labeled ``$(2)$''.

\subsection{Quantities Measured}
The main quantities measured in this simulation are the spin glass
susceptibility $\chi_{SG}$, and the chiral glass susceptibility $\chi_{CG}$,
at wavevectors $k=0$, and $k=2\pi/N$, and from these we obtain the two corresponding
correlation lengths, $\xi_{SG}$ and $\xi_{CG}$.  The spin
glass order parameter, $q^{\mu\nu}({k})$, at wave vector ${k}$, is defined to
be
\begin{equation}
q^{\mu\nu}({k}) = {1 \over N} \sum_{i=1}^N S_i^{\mu(1)} S_i^{\nu(2)}
e^{i {k} \cdot {R}_i},
\label{qmunu}
\end{equation}
where $\mu$ and $\nu$ are spin components, and ``$(1)$'' and ``$(2)$''
denote two
identical copies of the system with the same interactions. We run two copies
of the system at each temperature in order to evaluate quantities such as the
spin glass susceptibility, defined in Eq.~\eqref{chisgk} below, without bias.
From this we
determine the wave vector dependent
spin glass susceptibility $\chi_{SG}({k})$ by
\begin{equation}
\chi_{SG}({k}) = N \sum_{\mu,\nu} [\langle \left|q^{\mu\nu}({k})\right|^2 \rangle ]\av ,
\label{chisgk}
\end{equation}
where $\langle \cdots \rangle$ denotes a thermal average and
$[\cdots ]\av$ denotes an average over disorder. The spin glass correlation
length is then determined from
\begin{equation}
\xi_{SG} = {1 \over 2 \sin (k_\mathrm{min}/2)}
\left({\chi_{SG}(0) \over \chi_{SG}({k}_\mathrm{min})} - 1\right)^{1/(2\sigma-1)},
\label{chisg}
\end{equation}
where ${k}_\mathrm{min} = (2\pi/L)$.
For the Heisenberg spin glass, Kawamura
defines the local chirality in terms of three spins on a line as
follows~\cite{hukushima:00}:
\begin{equation}
\kappa_i = {\bf S}_{i+1} \cdot {\bf S}_i \times {\bf S}_{i-1}.
\label{eq:chiral_heis}
\end{equation}
The chiral glass susceptibility is then given by
\begin{equation}
\label{chicg}
\chi_{CG}({k}) =  N [\langle \left| q_{c}({k})\right|^2
\rangle ]\av ,
\end{equation}
where the chiral overlap $q_{c}({\bf k})$ is given by
\begin{equation}
\label{qc}
q_{c}({k}) = {1 \over N} \sum_{i=1}^N  \kappa_i^{(1)} \kappa_i^{(2)}
e^{i {k} \cdot {R}_i}.
\end{equation}
We define
the chiral correlation length by
\begin{equation}
\label{xi_c}
\xi_{CG} = {1 \over 2 \sin (k_\mathrm{min}/2)}
\left({\chi_{CG}(0) \over \chi_{CG}({k}_\mathrm{min})} - 1\right)^{1/(2\sigma-1)}.
\end{equation}
As will be revealed in the next section, three of the four quantities defined
above, $\chi_{SG},\xi_{SG},$ and $\chi_{CG}$ may be used in a
finite-size-scaling analysis to locate and analyze the phase transition.

\subsection{Finite-Size Scaling}
According to finite-size
scaling~\cite{fss:gtlcd}, the correlation length of the finite-system varies,
near the transition temperature $T_c$,
as
\begin{subequations}
\label{eq:xiscale}
\begin{align}
\label{xi_nonmf}
{\xi \over N} &= {\mathcal X} [ N^{1/\nu} (T - T_c) ]
%&& {\xi}/{N} &\sim {\mathcal X} [ N^{1/\nu} (T - T_c) ]
\; , \ (2/3 \le \sigma < 1), 
\\
{\xi \over N^{\nu/3}} &= {\mathcal X} [ N^{1/3} (T - T_c) ]
%&& {\xi_N}/{N^{\nu/3}} &\sim {\mathcal X} [ N^{1/3} (T - T_c) ]
\;,\ (1/2 <\sigma \le 2/3),
\label{xi_mf}
\end{align}
\end{subequations}
in which $\nu$, the correlation length exponent, is given, in the mean-field
regime, by $\nu = 1/(2\sigma - 1)$. We will use Eq.~\eqref{eq:xiscale}
for both the spin glass correlation length $\xi_{SG}$, in which $T_c$
will be set to $T_{SG}$, and the chiral glass correlation length $\xi_{CG}$,
in which $T_c$ will be set to $T_{CG}$. It follows that,
if there is a transition at $T = T_c$,
data for $ {\xi}/{N}$ ($ {\xi}/{N^{\nu/3}}$ in the mean-field region)
for different system sizes $N$ should cross at $T_c$.

We also present data for $\chi_{SG} \equiv \chi_{\rm SG}(0)$, which
has the finite-size scaling form
\begin{subequations}
\label{eq:chisgscale}
\begin{align}
\label{chi_nonmf}
{\chi_{\rm SG} \over N^{2 -\eta}} &= {\mathcal C}[N^{1/\nu} (T - T_c)]
\; , \ (2/3 \le \sigma < 1), 
\\
{\chi_{\rm SG} \over N^{1/3}} &= {\mathcal C}[N^{1/3} (T - T_c)]
\; , \ (1/2 < \sigma \le 2/3).
\label{chi_mf}
\end{align}
\end{subequations}
Hence curves of $\chi_{\rm SG}/N^{2 - \eta}$ ($\chi_{\rm SG} /
N^{1/3}$ in the mean-field regime) should also intersect. This is particularly
useful for long-range models since $\eta$ is given by the simple expression $2
- \eta = 2 \sigma - 1$ {\it exactly}. However, we do not know the exponent
corresponding to $\eta$ for the chiral glass susceptibility, so we will not
use this quantity in the finite-size scaling analysis.

In practice, there are corrections to this finite-size-scaling, so 
data for different sizes do not all intersect at the exactly the same temperature.
Including leading corrections to scaling, the intersection
temperature $T^{*}(N,2N)$ for sizes $N$ and $2N$ varies
as~\cite{binder:81b,ballesteros:96a,hasenbusch:08b,larson:10}
\begin{equation}
T^{*}(N,2N) = T_{c} + \frac{A}{N^{\lambda}},
\label{Tstar}
\end{equation}
where $A$ is the amplitude of the leading correction, and the 
exponent $\lambda$ is given by
\begin{equation}
\lambda = \frac{1}{\nu}+\omega
\label{lambda}
\end{equation}
where $\omega$ is the leading correction to scaling
exponent.

\section{Results}
\label{sec:results}
\begin{figure}
\begin{center}
\includegraphics[width=\columnwidth]{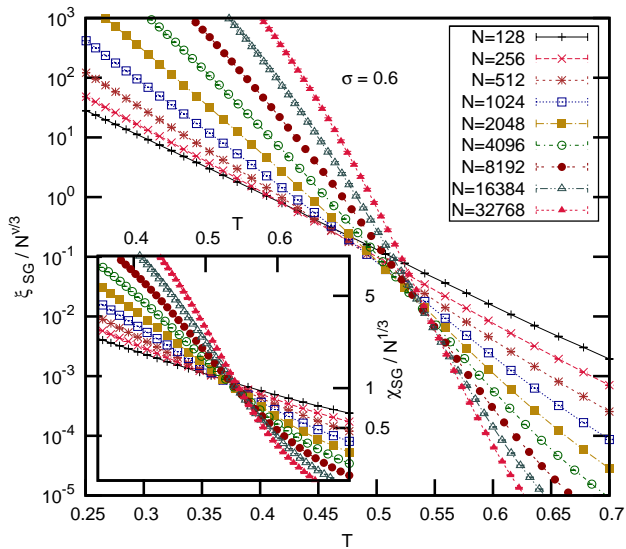}
\caption{(Color online)
The  main figure shows
data for $\xi_{SG}/N^{5/3}$, in which the power of $N$ is chosen following
Eq.~\eqref{xi_mf} with $\nu=1/(2\sigma-1)$,
as a function of $T$ for
different system sizes for $\sigma=0.6$. The inset shows data for 
$\chi_{SG}/N^{1/3}$, in which the power of $N$ is chosen following Eq.~\eqref{chi_mf}.
}
\label{fig:both_0.6}
\end{center}
\end{figure}

\begin{figure}
\includegraphics[width=\columnwidth]{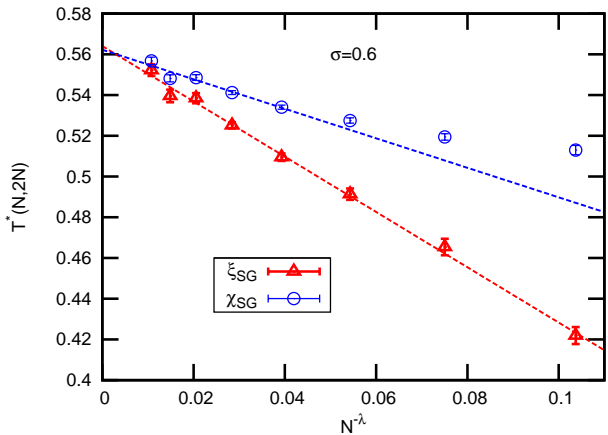}
\caption{(Color online)
The intersection temperatures, $T^{*}(N,2N)$, obtained from the data
in  Fig.~\ref{fig:both_0.6}, for
$\xi_{SG}/N^{5/3}$ and $\chi_{SG}/N^{1/3}$ for $\sigma = 0.6$,
as a
function of $N^{-\lambda}$, with $\lambda = 0.467$ determined from
Eq.~\eqref{lambda} (which is valid
in the MF regime). A fit using all 8 points from $\xi_{SG}$
gives $T_{SG} = 0.564 \pm 0.002$, while a fit using the data for the largest
5 pairs of sizes from $\chi_{SG}$ gives
$T_{SG} = 0.562 \pm 0.002$.
}
\label{fig:Tstar_0.6}
\end{figure}

%This section concerns with the presentation, and analysis of the key results
%of the simulation. We present finite-size-scaling figures for the quantities
%$\xi_{SG}$, $\xi_{CG}$ and $\chi_{SG}$ scaled by the appropriate powers of the
%system size obtained from standard finite-size-scaling theory. It is known
%that the quantities $\xi_{SG}/N$, $\xi_{CG}/N$, and $\chi_{SG}/N^{2-\eta}$,
%where $\eta$ is a critical exponent, are in finite-size-scaling form. 

\subsection{$\sigma = 0.6$ (mean-field regime)}
%The finite scaling data for $\sigma=0.6$ are discussed in this subsection.
As shown in Table~\ref{tab:ranges}, $\sigma=0.6$
lies well inside the mean-field regime. 
Hence, according to Eq.~\eqref{xi_mf}, results for
$\xi_{SG}/N^{\nu/3}$ should intersect at $T_{SG}$ with 
$\nu$ set equal to $1/(2\sigma-1)$.
The data is shown in the main part of
Fig.~\ref{fig:both_0.6}. The intersections do not occur at
precisely the same temperature, but fitting the intersection temperatures to
Eq.~\eqref{Tstar} is helped by the fact that we know 
$\lambda \equiv \frac{5}{3}-2\sigma$ in the MF regime~\cite{larson:10},
which gives a value $0.467$ here. 
A straight line fit of $\xi_{SG}/N^{\nu/3}$ against $N^{-\lambda}$,
shown in Fig.~\ref{fig:Tstar_0.6}, gives $T_{SG}=0.564 \pm 0.002$.

The inset to Fig.~\ref{fig:both_0.6}
shows data for $\chi_{SG}/N^{1/3}$, which should also intersect
at $T_{SG}$ according to
Eq.~\eqref{chi_mf}.
This time, we find that corrections to scaling are well described
Eq.~\eqref{Tstar}
but only if we consider just
the largest five pairs of sizes. The fit to Eq.~\eqref{Tstar},
shown in Fig.~\ref{fig:Tstar_0.6}, gives $T_{SG}=0.562 \pm 0.002$, which is
consistent with that obtained from the spin glass correlation
length.

We have also measured the chiral glass correlation length.
However, we find
that
$\xi_{CG}/N^{\nu/3} \lesssim 10^{-12}$ in the vicinity of $T_{SG}$.
Hence chiralities can not play an important role in this range of temperature.

\begin{figure}
\includegraphics[width=\figurewidth]{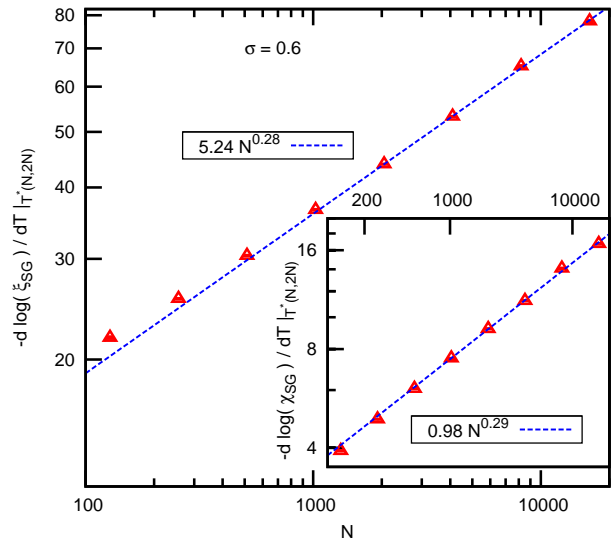}
\caption{(Color online)
The main figure is a log-log plot of the logarithmic derivative of $\xi_{SG}$
for $\sigma=0.6$
for different sizes $N$ evaluated at the intersection temperatures $T^\star(N,
2N)$ shown in Fig.~\ref{fig:Tstar_0.6}. According to
Eq.~\eqref{xi_mf} the slope is expected to be $1/3$. The
best fit is a little smaller than this, indicating that corrections to scaling are still
present for these sizes. The inset is the same but for $\chi_{SG}$.
}
\label{fig:both_deriv_0.6}
\end{figure}

According to Eqs.~\eqref{xi_mf} and \eqref{chi_mf} the argument of the scaling
functions is $N^{1/3}(T - T_c)$. Hence, at $T_{SG}$ the, logarithmic derivative
of $\xi_{SG}$ and $\chi_{SG}$ should vary as $N^{1/3}$. As we have seen, the
data do not all intersect at the same temperature, and so we evaluate the
derivatives at the intersection temperatures $T^\star$ plotted in
Fig.~\ref{fig:Tstar_0.6}. The results are shown in
Fig.~\ref{fig:both_deriv_0.6}. We get a power of 0.28 from $\xi_{SG}$ and 0.29
from $\chi_{SG}$,
in both cases a little less than
1/3, indicating that there are still some corrections to scaling even for
these large sizes.

\subsection{$\sigma = 0.75$ (non mean-field regime)}

\begin{figure}
\includegraphics[width=\figurewidth]{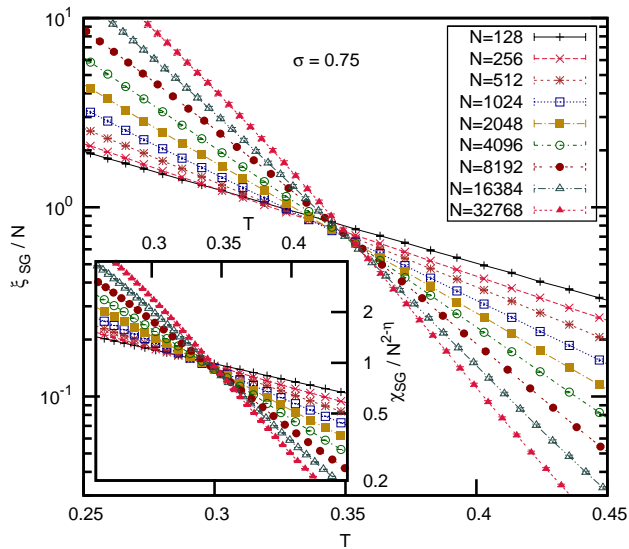}
\caption{(Color online)
The main figure shows data for different sizes for $\xi_{SG}/N$ at $\sigma =
0.75$ which,
according to Eq.~\eqref{xi_nonmf}, should intersect at $T_{SG}$. The inset
shows data for $\chi_{SG}/N^{2-\eta}$ (with $2-\eta = 2\sigma - 1$), which should
also intersect at $T_{SG}$ according to Eq.~\eqref{chi_nonmf}.
}
\label{fig:both_0.75}
\end{figure}
\begin{figure}
\includegraphics[width=\columnwidth]{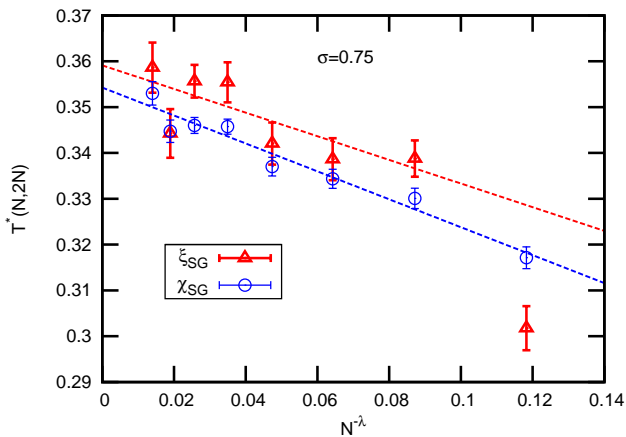}
\caption{(Color online)
The intersection temperatures for $\sigma = 0.75$. The all 8 data points
from $\chi_{SG}$ were fitted
to Eq.~\eqref{Tstar}, with the result that $\lambda = 0.44 \pm 0.13$. The same
exponent was then used to fit the results for
the largest 7 (pairs of) sizes from
$\xi_{SG}$, for which sub-leading
corrections to scaling seem to be more important. The resulting values for the
transition temperature are $T_{SG} = 0.354 \pm 0.005$ 
from $\chi_{SG}$, and
$T_{SG} = 0.359 \pm 0.003$ from $\xi_{SG}$.
}
\label{fig:Tstar_0.75}
\end{figure}

For $\sigma = 0.75$ we are no longer in the MF regime.
Hence,  according to Eqs.~\eqref{xi_nonmf} and
\eqref{chi_nonmf}, $\xi_{SG}/N$
and $\chi_{SG}/N^{2-\eta}$ (with $2-\eta = 2 \sigma - 1$), should
intersect at $T_{SG}$. The data
are shown in Fig.~\ref{fig:both_0.75} and the resulting intersection
temperatures are shown in Fig.~\ref{fig:Tstar_0.75}.

We fit the intersection temperatures to Eq.~\eqref{Tstar}, but unfortunately
we do not know the value of the exponent $\lambda$ outside the MF region, and
have to include it as fit parameter.
The fit is therefore to
a non-linear function of the parameters. We determine the fit
parameters 
using the Levenberg-Marquardt algorithm~\cite{press:92}.
The data from
$\chi_{SG}$ is better behaved than the data from $\xi_{SG}$ so we use the
former to determine the exponent $\lambda$ and then fix this value in the fit
(which did not include the smallest size)
to the data from $\xi_{SG}$. This procedure is justified since the exponent
giving the leading correction to scaling, $\lambda$, is universal, though the
amplitude of this correction ($A$ in Eq.~\eqref{Tstar}) is non-universal.
The results are $\lambda = 0.44 \pm 0.13$,
$T_{SG} = 0.359 \pm 0.003$ from $\xi_{SG}$, and $T_{SG} = 0.354 \pm 0.005$
from $\chi_{SG}$. The two estimates for $T_{SG}$ agree within the error bars.

The data for $\xi_{CG}/N$ in the region of the spin glass transition
temperature is very small, around $10^{-4}$. Hence, as for $\sigma = 0.6$,
chiralities do not play an important role in the transition.

\begin{figure}
\includegraphics[width=\figurewidth]{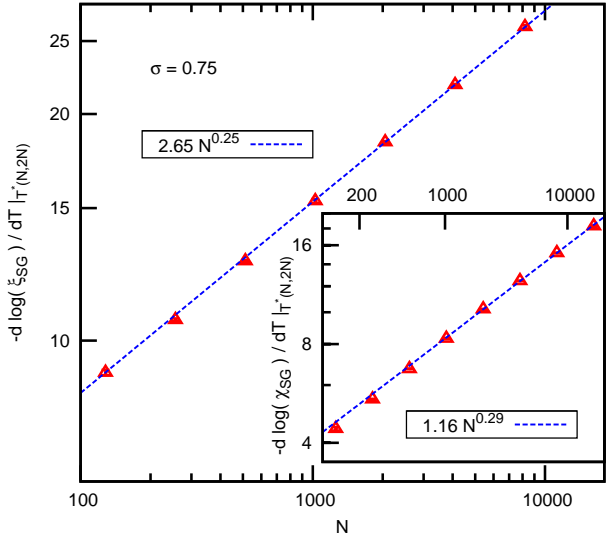}
\caption{(Color online)
The main figure is a log-log plot of the logarithmic derivative of $\xi_{SG}$
for $\sigma = 0.75$
for different sizes $N$ evaluated at the intersection temperatures $T^\star(N,
2N)$ shown in Fig.~\ref{fig:Tstar_0.75}. According to
Eq.~\eqref{xi_nonmf} the slope is expected to be $1/\nu$.
The inset is the same but for $\chi_{SG}$.
}
\label{fig:both_deriv_0.75}
\end{figure}

According to Eqs.~\eqref{xi_nonmf} and \eqref{chi_nonmf}, adapted to include 
corrections to finite-size scaling, the logarithmic
derivative of $\xi_{SG}$ and $\chi_{SG}$ should vary as $N^{1/\nu}$ at
$T^\star(N, 2N)$. The plots in Fig.~\ref{fig:both_deriv_0.75} yield
$1/\nu_{SG} = 0.25$ from $\xi_{SG}$ and $0.29$ from $\chi_{SG}$. The
difference presumably comes from corrections to scaling.

Kotliar et al.~\cite{kotliar:83} calculated critical exponents to leading 
order in $\epsilon = \sigma - 2/3$ with the result 
\begin{equation}
{1 \over \nu} = {1 \over3}\, \left(1 - 12 \epsilon + O(\epsilon^2) \right) \, .
\label{eps_exp}
\end{equation}
The large coefficient of $\epsilon$ indicates the expansion becomes poorly
converged well before the ``lower critical'' value $\epsilon = 1/3$ ($\sigma =
1$). Even for the present value of $\sigma$,
which corresponds to $\epsilon= 1/12$, Eq.~\eqref{eps_exp} gives $1/\nu = 0$,
and so is not useful for comparison with the numerics.

\subsection{$\sigma = 0.85$ (non mean-field regime)}

\begin{figure}
\begin{center}
\includegraphics[width=\figurewidth]{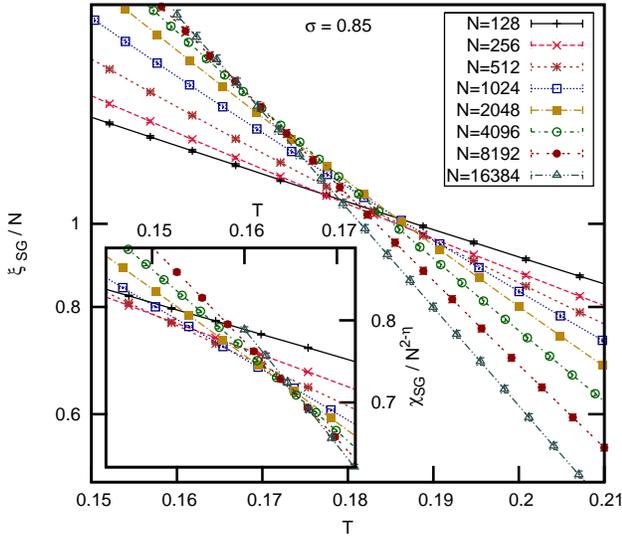}
\caption{(Color online)
The main figure shows data for $\xi_{SG}/N$,
as a function of $T$ for different system sizes for $\sigma=0.85$.
The inset shows data for $\chi_{SG}/N^{2-\eta}$ with $2-\eta = 2\sigma - 1$.
}
\label{fig:both_0.85}
\end{center}
\end{figure}

\begin{figure}
\includegraphics[width=\figurewidth]{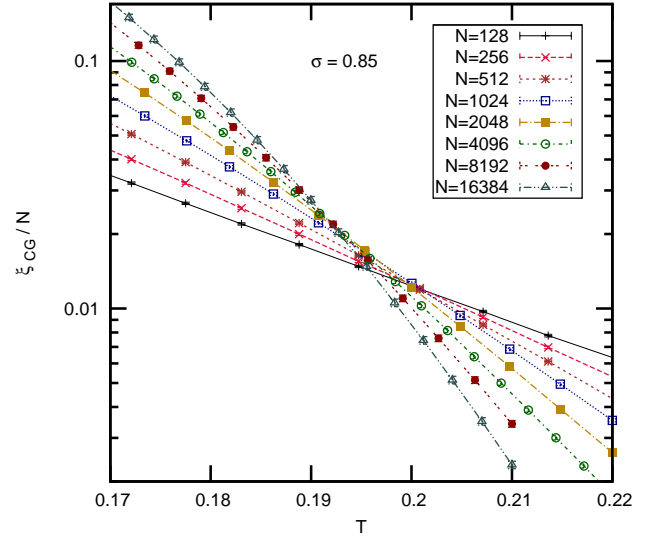}
\caption{(Color online)
Data for $\xi_{CG}/N$, the spin glass correlation length divided by system
size, as a function of $T$ for different system sizes for $\sigma=0.85$.
}
\label{fig:xi_CG_0.85}
\end{figure}

\begin{figure}
\includegraphics[width=\columnwidth]{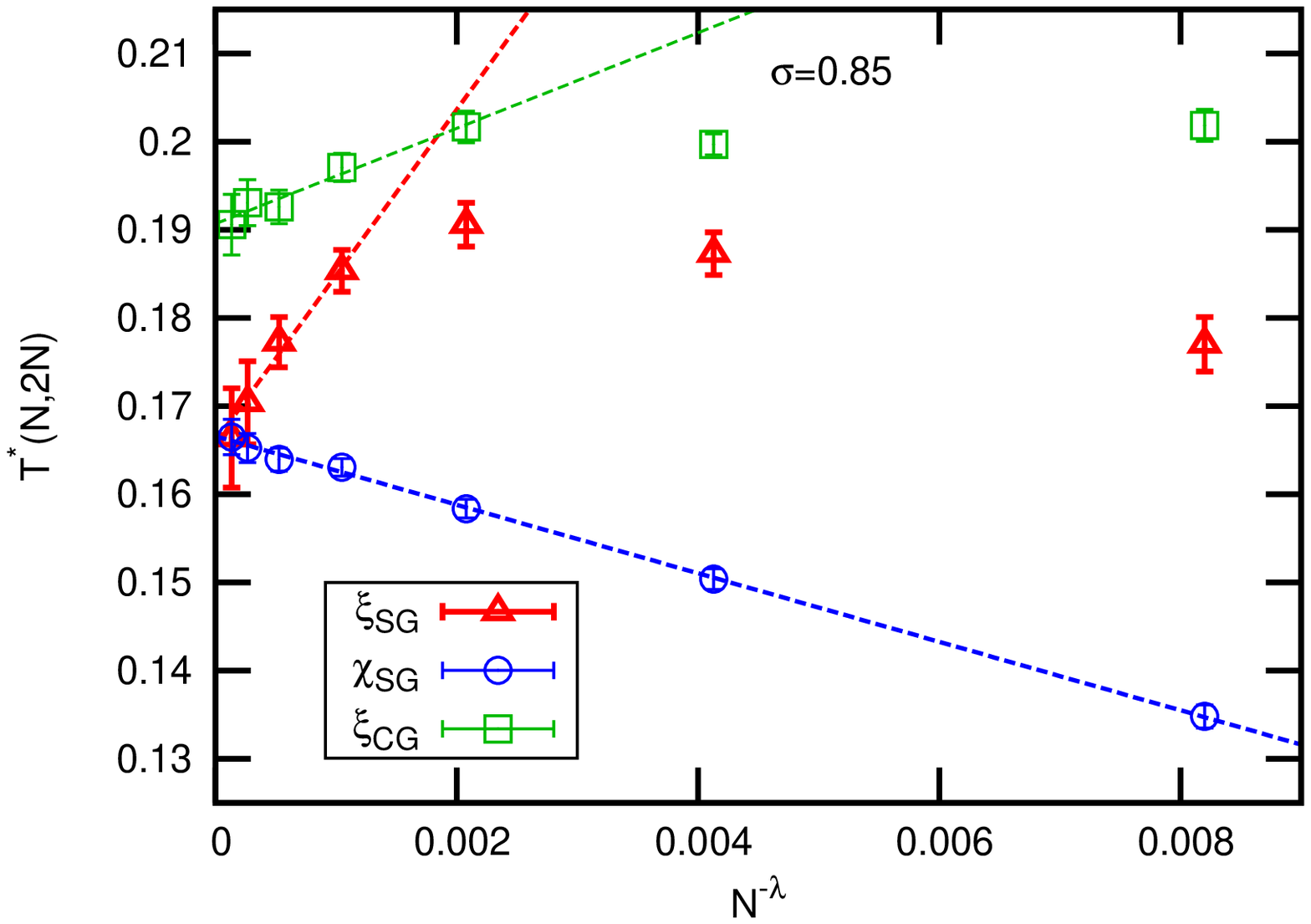}
\caption{(Color online)
The intersection temperatures for $\sigma = 0.85$. The data from $\chi_{SG}$
was fitted
to Eq.~\eqref{Tstar}, with the result that $\lambda = 0.99 \pm 0.13$. This is
the value used to plot all the data.
The same
exponent was then used to fit the largest four sizes in the
data from $\xi_{SG}$, for which sub-leading corrections appear to be very
large.
The full Levenberg-Marquardt fit to the data for $\xi_{CG}$ was unable to determine
$\lambda$ with any precision, see text.
%and so, in the plot we also set $\lambda = 0.99$. 
However, in estimating $T_{CG}$ and its error
bar from data for $\xi_{CG}$ we 
allowed $\lambda$ to vary.
The resulting values for the
transition temperatures are
$T_{SG} = 0.167 \pm 0.001$ from $\chi_{SG}$ (all 7 points),
$T_{SG} = 0.166 \pm 0.004$ from $\xi_{SG}$ (4 points), and
$T_{CG} = 0.190 \pm 0.006$ from $\xi_{CG}$ (5 points).
We emphasize that, in these estimates, we fixed the
value of $\lambda$ only for the data from $\xi_{SG}$. 
}
\label{fig:Tstar_0.85}
\end{figure}

For $\sigma = 0.85$ we are further in the non-mean-field region. Among the
different models studied here, this is the one which is most similar to
a short-range model in three dimensions.
The spin
glass data is shown in Fig.~\ref{fig:both_0.85}. Chiral correlations are
larger than for $\sigma
= 0.6$ and $0.75$, so we show data
for $\xi_{CG}$ in Fig.~\ref{fig:xi_CG_0.85}.

As for $\sigma = 0.75$, to extrapolate the intersection temperatures to the
thermodynamic limit, we resort to Levenberg-Marquardt fits with
three parameters. Using data from $\chi_{SG}$ we find $\lambda_{SG} = 0.99 \pm
0.13$. The data
from $\xi_{CG}$ is insufficient to determine $\lambda_{CG}$ since the fits give
$\lambda_{CG} = 0.79 \pm 0.74$, i.e. the error bar is as large as the best
estimate. Nonetheless, in estimating $T_{CG}$ and its error bar from the data for
$\chi_{CG}$, we allow $\lambda$ to vary. For $\xi_{SG}$ there
appear to be very large sub-leading corrections, so
we fixed the value of $\lambda_{SG}$ to that obtained from $\chi_{SG}$ when
fitting the results from $\xi_{SG}$.

Results for $T^\star(N, 2N)$ against $1/N^\lambda$ and fits are shown
in Fig.~\ref{fig:Tstar_0.85}. In the \textit{plot}, for all data
we use the value of $\lambda$ determined from $\chi_{SG}$. However,
we again emphasize that,
in the
\textit{fit} to the $\chi_{CG}$ data, we estimated  $T_{CG}$ and its error bar by
allowing $\lambda$ to vary.
%Since $\lambda_{CG}$ is so poorly determined, in
%the plot, we
%use the value obtained from $\chi_{SG}$.
From the fits, we
find $T_{SG} = 0.167 \pm 0.001 $ from $\chi_{SG}$,
$T_{SG} = 0.166 \pm 0.004$ from $\xi_{SG}$, and
$T_{CG} = 0.190 \pm 0.006$ from $\xi_{CG}$.
%We emphasize that while the value
%of $T_{SG}$ and its error bar determined from $\xi_{SG}$ were determined with
%a fixed value of $\lambda$, the value of $T_{CG}$ and its error bar determined
%from $\chi_{SG}$ were obtained allowing $\lambda$ to vary.

The two estimates of $T_{SG}$ agree
with each other but are lower than $T_{CG}$. This would imply spin-chirality
decoupling for $\sigma = 0.85$. However, we note that the data for the spin
glass correlation length appears, at intermediate sizes, to be 
extrapolating to a value for
$T_{SG}$ of around 0.19 (our value for $T_{CG}$) but then,
for the largest sizes, veers down to about
$0.167$ (very close to our value for $T_{SG}$ obtained from $\chi_{SG}$).
Hence we cannot rule out the possibility that a similar
``crossover'' may occur for the chiral glass correlation length data, but at
even larger sizes. If so, then spin-chirality decoupling would not occur at
the largest scales. We also note that the actual values of 
$\xi_{CG}/N$ shown in Fig.~\ref{fig:xi_CG_0.85} are still very small in the
vicinity of
$T_{CG}$, about 1/50 of
the value of $\xi_{SG}/N$ around the transition, 
see Fig.~\ref{fig:both_0.85}. Hence we are very far the regime with $\xi_{CG}
> \xi_{SG}$ which will ultimately occur in the presence of spin-chirality
decoupling. 

\begin{figure}
\includegraphics[width=\figurewidth]{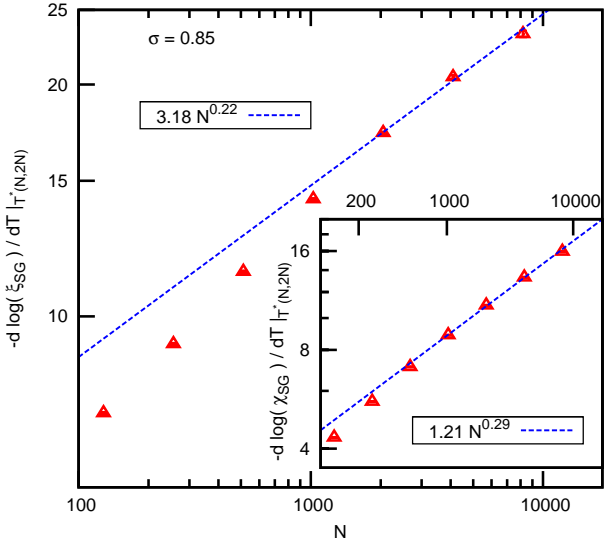}
\caption{(Color online)
The main figure is a log-log plot of the logarithmic derivative of $\xi_{SG}$
for $\sigma = 0.85$
for different sizes $N$
evaluated at the intersection temperatures $T^\star(N,
2N)$ shown in Fig.~\ref{fig:Tstar_0.85}. According to
Eq.~\eqref{xi_nonmf} the slope is expected to be $1/\nu$.
The inset is the same but for $\chi_{SG}$.
}
\label{fig:both_deriv_0.85}
\end{figure}

\begin{figure}
\includegraphics[width=\figurewidth]{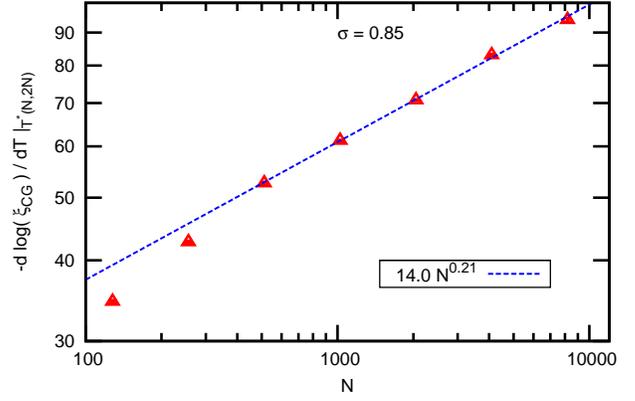}
\caption{(Color online)
Similar to Fig.~\ref{fig:both_deriv_0.85} but for $\xi_{CG}$.
}
\label{fig:xi_CG_deriv_0.85}
\end{figure}

Figure \ref{fig:both_deriv_0.85} shows results for the logarithmic derivative
of $\xi_{SG}$ and $\chi_{SG}$ evaluated at $T^\star(N, 2N)$. Fits give
$1/\nu$ = 0.22 from $\xi_{SG}$ and 0.29 from $\chi_{SG}$.
The curvature in the data for $\xi_{SG}$ indicates strong
finite-size corrections. Presumably these
corrections are also the reason for the
difference between the two estimates for $1/\nu$.
Figure \ref{fig:xi_CG_deriv_0.85} shows similar data but for $\xi_{CG}$. The
best fit gives $1/\nu = 0.21$, which is close to the estimate from $\xi_{SG}$.
Note that the coefficient of $N^{1/\nu}$,
$14.01$, is much larger than the corresponding value,
3.18, for $\xi_{SG}$, presumably to compensate for the overall value of
$\xi_{CG}$ being much less than that of $\xi_{SG}$ in the vicinity of the
intersection temperatures $T^\star$.


\subsection{$\sigma = 1\quad (= \sigma_l)$}

\begin{figure}
\includegraphics[width=\figurewidth]{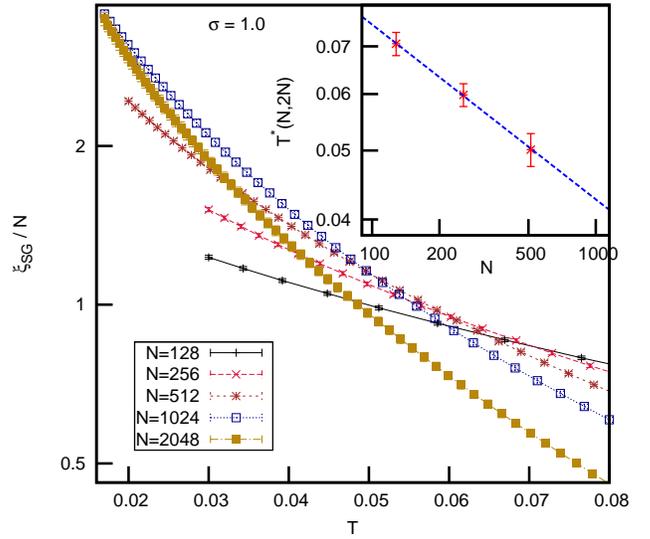}
\caption{(Color online)
Data for $\xi_{SG}/N$, the spin glass correlation length divided by system
size, as a function of $T$ for different system sizes, for $\sigma=1.0$. The
inset shows the intersection temperatures $T^\star(N, 2N)$, as well as
a log-log fit assuming $T_{SG}=0$. The fit works well for $N=128,256$ and $512$.
There is no intersection
for the two lowest sizes, $N=1024$ and $2048$ for $T$ greater than the lowest
temperature we could simulate, $0.017$. This temperature is well
\textit{below} the value of the fit extrapolated to $N=1024$, which is about
0.042. Hence the intersection temperatures actually fall off \textit{faster}
at large sizes than shown in the fit.
}
\label{fig:corrsg1}
\end{figure}

\begin{figure}
\includegraphics[width=\figurewidth]{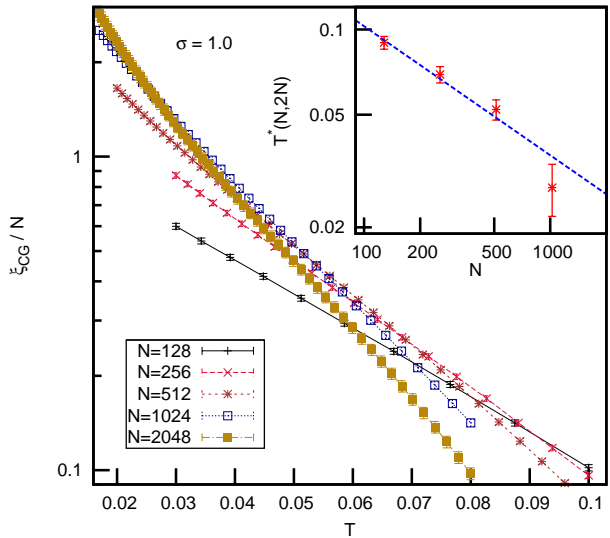}
\caption{(Color online)
Data for $\xi_{CG}/N$, the chiral glass correlation length divided by system
size, as a function of $T$ for different system sizes, for $\sigma=1.0$. The
inset is a log-log fit to the intersection temperatures
$T^\star(N, 2N)$, assuming $T_{CG}=0$. The data is quite consistent with this
behavior.
%In fact, it seems to suggest that a
%negative $T_{c}$ which is unphysical, would give a better fit.
}
\label{fig:corrcg1}
\end{figure}
\begin{figure}
\includegraphics[width=\figurewidth]{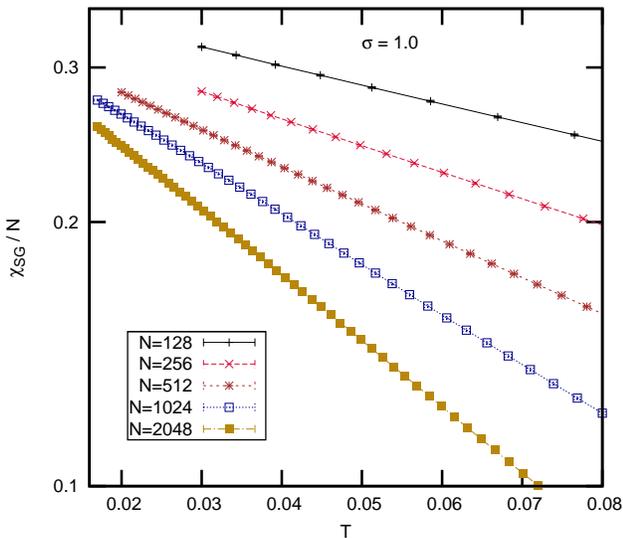}
\caption{(Color online)
Data for $\chi_{SG}/N$, the spin glass susceptibility divided by system size
(the power $2-\eta$ is equal to 1 here), as a function of $T$ for different
system sizes for $\sigma=1.0$. There are no intersections for the
temperature-range in which the simulation is conducted.
%However, it seems very
%plausible that the curves corresponding to different sizes in the figure will
%intersect at zero temperature. 
}
\label{fig:khisg1}
\end{figure}

It is known from the early work of Kotliar et al.~\cite{kotliar:83} that
$\sigma = 1$ is the ``lower critical'' value $\sigma_l$, above which there is no spin
glass transition. Interestingly, Viet and
Kawamura~\cite{viet:10} claim that chiral
glass ordering persists to slightly \textit{larger} values of $\sigma$. 
Testing this claim is one of our main motivations for performing
simulations at $\sigma = 1$.

Figure \ref{fig:corrsg1} shows the finite scaling for $\xi_{SG}$. In the inset,
we show a log-log plot of $T^{*}(N,2N)$ versus $N$ and include a straight-line fit for
$N = 128, 256$ and $512$. This fit works well. We find no intersections in the
range of $T$ that we can equilibrate $(T \ge 0.17)$ for $N=1024$ and 2048.
Hence the data is well consistent with $T_{SG} =0$.

Figure \ref{fig:corrcg1} shows the corresponding figure for
$\xi_{CG}$. Again, the inset shows log-log fits assuming that the transition 
temperature, $T_{CG}$ in this case, is zero.
The fit is satisfactory and indicates that we do not find a
finite value for $T_{CG}$ at $\sigma = 1$, in contrast to the claim of
Viet and Kawamura~\cite{viet:10}.

Figure \ref{fig:khisg1} shows the data for $\chi_{SG}$.  There are no
intersections at all in the range of temperature that we can equilibrate,
consistent with the conclusion from the $\xi_{SG}$ data that $T_{SG} =
0$.

\section{Conclusions}
\label{sec:conclusions}

Our primary motivation to study
the Heisenberg spin glass in one dimension with long-range
interactions which fall off with the power of the distance, is to test 
Kawamura's
spin-chirality decoupling scenario in which $T_{CG} > T_{SG}$, and his
subsequent prediction~\cite{viet:10}
that chiral glass ordering persists for $\sigma >
\sigma_l$, where $\sigma_l = 1$ is the ``lower critical'' value for the spin
glass transition, with a finite value of $T_{CG}$ \textit{at} $\sigma = 1$. 

For $\sigma = 1$ we find $T_{CG} = T_{SG} =0$ in contrast to Viet and
Kawamura~\cite{viet:10}. For most of the other values of $\sigma$
our data is well consistent with a single phase transition.

However, for $\sigma = 0.85$ the
best fits for the sizes we can equilibrate indicate $T_{CG} > T_{SG}$, see
Fig.~\ref{fig:Tstar_0.85}. Interestingly that figure shows very strong
subleading corrections to finite-size scaling in the data for $\xi_{SG}$ since
it only fits Eq.~\eqref{Tstar} for the largest sizes. At
intermediate sizes the intersection temperatures seem to be heading towards
the chiral glass transition temperature obtained from the $\xi_{SG}$ data,
but then dip down for the largest
sizes. We therefore cannot rule out that similar behavior
might occur for the chiral glass correlation length but at even larger length
scales. In this case there would be no spin-chirality decoupling.  We also
note that, for a given size and temperature, $\xi_{CG}$ remains considerably
smaller than $\xi_{SG}$ in the vicinity of the intersection temperatures
$T^\star$, compare the main part of Fig.~\ref{fig:both_0.85}
with Fig.~\ref{fig:xi_CG_0.85}.
Hence the data is very far from the regime with
$\xi_{CG} > \xi_{SG}$ which ultimately prevails for $T_{SG} < T< T_{CG}$
if there is spin-chirality decoupling.

In a subsequent paper~\cite{sharma:11b}, we will investigate,
for the same models, under what
circumstances an AT line of transitions occurs in a magnetic field.

\acknowledgments
We acknowledge support from the NSF under Grant DMR-0906366. AS also acknowledges
partial support from DOE under Grant No.~FG02-06ER46319. We are grateful
for a generous allocation of computer time from the Hierarchical Systems
Research Foundation.

\bibliography{refs,comments}

\begin{thebibliography}{47}
\bibitem[{fss()}]{fss:gtlcd}
\bibinfo{note}{For a discussion of how finite-size scaling is modified in the
  region of mean-field exponents, see for example
  Refs.~[\onlinecite{brezin:82,brezin:85,binder:85,luijten:99,jones:05}].}
\end{thebibliography}

\end{document}